\documentclass[useAMS,usenatbib,usegraphicx,usedcolumn]{mn2e}

\title[Instability of Pop III Very Massive Stars]{Vibrational Instability of Population III Very Massive Main-Sequence Stars due to the $\varepsilon$-Mechanism}
\author[T. Sonoi and H. Umeda]{Takafumi Sonoi$^{1}$\thanks{E-mail:
sonoi@astron.s.u-tokyo.ac.jp} and Hideyuki Umeda$^{1}$ \\
$^{1}$Department of Astronomy, Graduate School of Science, The University of Tokyo, Hongo 7-3-1, Bunkyo-ku, Tokyo, Japan 113-0033}
\begin{document}

\date{}

\pagerange{\pageref{firstpage}--\pageref{lastpage}} \pubyear{2002}

\maketitle

\label{firstpage}

\begin{abstract}
Very massive stars are thought to be formed in the early Universe because of a lack of cooling process by heavy elements, and might have been responsible for the later evolution of the Universe. We had an interest in vibrational stability of their evolution and carried out the linear nonadiabatic analysis of radial and nonradial oscillations for population III very massive main-sequence stars with $500-3000M_{\sun}$. We found that only the radial fundamental mode becomes unstable due to the $\varepsilon$-mechanism for these stars. The instability appears just after the CNO cycle is activated and the nuclear energy generation rate becomes large enough to stop the pre--main-sequence contraction, and continues during the early stage of the core hydrogen burning. Besides, we roughly estimated amount of mass loss due to the instability to know its significance.     
\end{abstract}


\section{Introduction}

The first-generation stars born in the early Universe are thought to have no, or few heavy elements. They might play an important role for chemical evolution of the early Universe. The initial mass function for the first generation stars has not been definitively determined yet. However, a lack of heavy elements leads to a deficiency of the coolants in the star-formation stage and hence to an expectation that very massive stars might have been formed \citep[e.g.][]{Bromm1999,Abel2002,Omukai2003}. 

\citet{Ledoux1941} suggested that very massive stars might experience vibrational instability in the core hydrogen-burning stage. In case of very massive stars, influence of radiation pressure makes the effective ratio of specific heat, $\gamma$, close to 4/3. Then pulsation amplitude in the core becomes comparable with that in the envelope. This situation is favorable for instability due to the $\varepsilon$-mechanism since excitation by this mechanism in the core can exceed flux dissipation in the envelope. Many authors have analyzed the stability of very massive stars and proposed the critical mass above which the instability occurs. With the pure electron scattering opacity the radial fundamental mode becomes unstable in case of $\ga 60M_{\sun}$ \citep{Schwarzschild1959}, while with the recent OPAL opacity table \citep{Rogers1992} the critical mass is $121M_{\sun}$ for the solar metallicity \citep{Stothers1992}. In the latter case, the fundamental mode is excited mainly by the $\kappa$-mechanism of the metallic absorption rather than the $\varepsilon$-mechanism.

In the zero-metallicity case, however, the metallic absorption never occurs. \citet{Baraffe2001} analyzed the stability of population III very massive stars and found that the fundamental mode becomes unstable due to the $\varepsilon$-mechanism for stars with $120-500M_{\sun}$. 

Population III very massive stars might have released a significant amount of heavy elements by supernova explosions and contributed to the chemical evolution of the early Universe. The stars with $\ga 300M_{\sun}$ experience the core-collapse supernova explosion at the end of their life, while those with $130-300M_{\sun}$ the pair-instability supernova explosion \citep{Ohkubo2006}. In addition, the former stars leave intermediate mass black holes, for which several candidates have been found as ultraluminous X-ray sources \citep{Feng2011}. Then, the evolution of the very massive stars is very important for the above phenomena, and have been investigated by several groups \citep[e.g.][]{Klapp1983, Klapp1984, Heger2003, Marigo2003, Ohkubo2006, Bahena2010}. In this paper, we focused on the vibrational stability of their evolution, and extended the \citeauthor{Baraffe2001}'s analysis toward the more massive stars. That is, we constructed the stellar models with $\geq 500M_{\sun}$ and carried out the linear nonadiabatic analysis against radial and nonradial oscillations. 

\section{Equilibrium models}
We adopted a Henyey-type code developed by \citet{Umeda2002,Umeda2005} and \citet{Ohkubo2006} to calculate the stellar evolution. We constructed zero-metallicity models with 500, 1000 and 3000 $M_{\sun}$ during the core hydrogen-burning stage. These stars were calculated without considering mass loss. {\it Overshooting is not taken into account in this study.}

For the zero-metallicity stars, the only possible way of starting hydrogen burning is the pp-chain no matter how massive the star is. Hence, the temperature near the stellar centre is much higher than in the case of metal-riched stars. During the pre--main-sequence stage, gravitational contraction releases energy so that the energy equilibrium in the whole star can be kept. When the central temperature reaches $\sim 10^7$ K, the pp-chain is activated and becomes to release enough energy to stop the gravitational contraction for stars with $\la 20M_{\sun}$. For the more massive stars, however, the gravitational contraction cannot be stopped only by the pp-chain burning \citep{Marigo2001}. When the central temperature reaches $\sim 10^8$ K, the triple alpha reaction produces enough $^{12}$C to activate the CNO cycle and then the contraction stops. Thus, the main energy source for core hydrogen-burning of population III massive stars is the CNO cycle. Table \ref{tab:1} shows property of the equilibrium models for the ZAMS stage, which we define as the time when the gravitational contraction stops.
The central temperature is hardly different among the different mass models and correponds to the occurence of the triple alpha reaction. Due to the large nuclear energy generation, the convective core is very large and occupies $\simeq$ 90 per cent of the stellar mass.

\begin{table}
  \caption{Property of zero-metallicity ZAMS models. \label{tab:1}}
  \begin{tabular}{cccccc}\hline\hline
    $M$  & $R$ & $\log T_{\rm eff}$ & $\log L$ & $\log T_{\rm c}$ & $M_{\rm cc}$ \\ 
    $(M_{\sun})$ & $(R_{\sun})$ & (K) & $(L_{\sun})$ & (K) & $(M_{\sun})$ \\ \hline
    500  & 10.4 & 5.04 & 7.10 & 8.17 & 445 \\ 
    1000 & 14.7 & 5.06 & 7.46 & 8.19 & 898 \\
    3000 & 25.6 & 5.07 & 7.99 & 8.21 & 2731 \\ \hline
  \end{tabular}

  \medskip
  $M_{\rm cc}$ denotes convective core mass.
\end{table}

\section{Linear nonadiabatic analysis of radial/nonradial oscillations}

We carried out a linear nonadiabatic analysis of radial ($l=0$) and nonradial ($l=1,\; 2$) oscillations. Any eigenfunction of linear oscillations of stars is expressed in terms of a combination of a spatitial function and a time varying function. The latter is expressed by $\exp(i\sigma t)$, where $\sigma$ denotes the eigenfrequency. The former, the spatial part, is a radial function for the radial oscillations, while decomposed into a spherical harmonic function, which is a function of the colatitude and the azimuthal angle, and a radial function for the nonradial oscillations. 

For the radial (non-radial) oscillations, equations governing the radial functions lead to a set of fourth-order (sixth-order) differential equation, of which coefficients are complex, including terms of frequency $\sigma$. Together with proper boundary conditions, this set of equations forms a complex eigenvalue problem with an eigenvalue $\sigma$. The real part of $\sigma$, $\sigma_{\rm R}$, represents the oscillation frequency, and the imaginary part of $\sigma$, $\sigma_{\rm I}$, gives the growth rate or the damping rate, depending on its sign. We used a Henyey-type code developed by \citet{Sonoi2012} to solve this eigenvalue problem. Note that for stability analysis related to the $\varepsilon$-mechanism, it is important to evaluate the temperature and density dependences of the nuclear reactions in the oscillation time scale. The detail is introduced by \citet{Sonoi2012}.   

We adopted the ``frozen-in convection'' approximation, in which perturbations of the convective flux are neglected. The very massive stars have the large convective core and the convection-pulsation interaction might affect the vibrational stability. But the uncertainty in the physical process remains still now, and hence we neglected it for simplicity.

\section{Results}
\subsection{Variation of stability with stellar evolution}
\label{sec:4.1}

\begin{figure}
  \includegraphics[width=84mm]{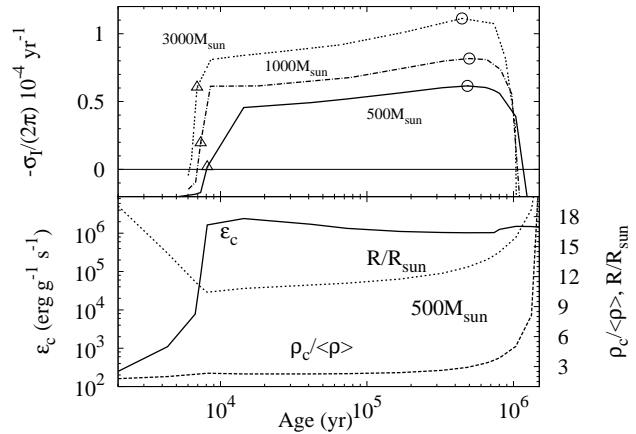}
  \caption{Top: variation of the growth rate of the radial fundamental mode for the 500, 1000 and 3000$M_{\sun}$ stars. The open triangles correspond to the ZAMS models, shown in table \ref{tab:1}, while the open circles correspond to the most unstable ones, for which the property of oscillations is shown in table \ref{tab:2}.  Bottom: variation of the nuclear energy generation rate at the centre $\varepsilon_{\rm c}$ and the ratio of the central density to the average density in the whole star $\rho_{\rm c}/\langle\rho\rangle$ for 500$M_{\sun}$.}
  \label{fig:1}
\end{figure}

\begin{table*}
  \begin{center}
    \caption{Property of oscillations for the models marked with the open circles in figure \ref{fig:1}. \label{tab:2}}
    \begin{tabular}{cccccccc} \hline
          &      & \multicolumn{3}{c}{Periods (h)}                & \multicolumn{3}{c}{Growth time-scale $\tau_{\rm g}$ (yr)} \\ 
      $l$ & Mode & $500M_{\sun}$ & $1000M_{\sun}$ & $3000M_{\sun}$ & $500M_{\sun}$      & $1000M_{\sun}$      & $3000M_{\sun}$ \\ \hline
      0   & F    & 4.31         & 6.17          & 10.4          & $1.63\times 10^4$ & $1.22 \times 10^4$ & $8.99 \times 10^3$ \\
          & 1H   & 1.70         & 2.13          & 2.98          & $-6.81\times 10$  & $-2.90\times 10$    & $-1.97\times 10$ \\
          & 2H   & 1.24         & 1.59          & 2.28          & $-2.35$           & $-7.12\times 10^{-1}$& $-5.45\times 10^{-1}$  \\
          &      &              &               &               &                   &                     &                   \\
      1   & g$_1$& 10.3         & 11.1          & 12.7          & $-9.20\times 10^4$ & $-1.54\times 10^5$ & $-3.51\times 10^5$ \\
          & p$_1$& 1.84         & 2.17          & 2.71          & $-4.42\times 10^2$ & $-8.31\times 10^2$ & $-7.37\times 10^2$ \\ 
          &      &              &               &               &                   &                     &                   \\
      2   & g$_1$& 6.32         & 6.77          & 7.64          & $-6.31\times 10^4$ & $-1.02\times 10^5$ & $-2.09\times 10^5$ \\
          & f    & 2.13         & 2.53          & 3.19          & $-2.40\times 10^3$ & $-3.55\times 10^3$ & $-4.46\times 10^3$ \\
          & p$_1$& 1.53         & 1.80          & 2.37          & $-8.21\times 10$   & $-1.88\times 10^2$ & $-2.13\times 10^2$ \\ \hline
    \end{tabular}
  \end{center}
\end{table*}

As described above, the gravitational contraction continues while only the pp-chain works. When the CNO cycle is activated, the contraction stops since the energy generation of the CNO cycle is much higher than the pp-chain and can maintain the energy equilibrium without the contraction.

Such high energy generation induces the vibrational instability due to the $\varepsilon$-mechanism; the nuclear energy generation increases at shrinking phase in pulsations. This makes expansion in the pulsations stronger and the nuclear energy generation drops more. This cyclical behaviour makes the star act as a heat engine and the pulsation amplitude will grow. 

Fig. \ref{fig:1} shows the variation of the growth rate, $-\sigma_{\rm I}/(2\pi)$, of the radial fundamental mode for 500, 1000 and 3000$M_{\sun}$ stars and the profiles of the equilibrium models for 500$M_{\sun}$ star with the stellar evolution. The positive value of the growth rate means instability, while the negative value stability. As the nuclear energy generation abruptly increases just before the ZAMS stage, the growth rate does as well and becomes positive. 

\begin{figure}
  \includegraphics[width=84mm]{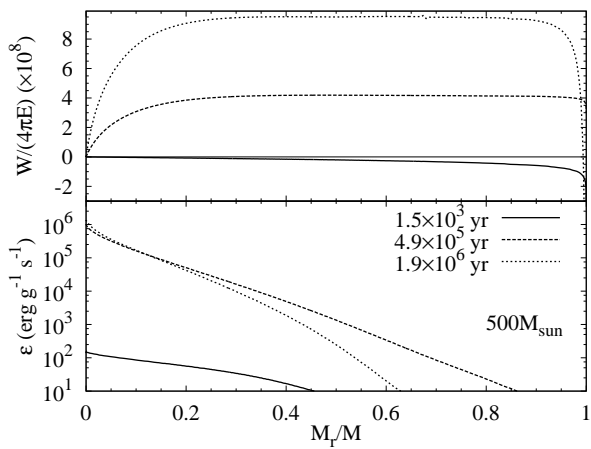}
  \caption{Workintegral $W$ for the radial fundamental mode (top) and nuclear energy generation rate $\varepsilon$ (bottom) for 500$M_{\sun}$. The workintegral is normalized with the oscillation energy of the mode in the whole star $E$.}
  \label{fig:2}
\end{figure}

Fig. \ref{fig:2} shows the work integral for the fundamental mode and the nuclear energy generation rate for 500$M_{\sun}$. Before the abrupt increase in the nuclear energy generation rate, the $\varepsilon$-mechanism does not work efficiently in the core. But after the increase, excitation by the $\varepsilon$-mechanism causes the instability. 

\begin{figure}
  \includegraphics[width=84mm]{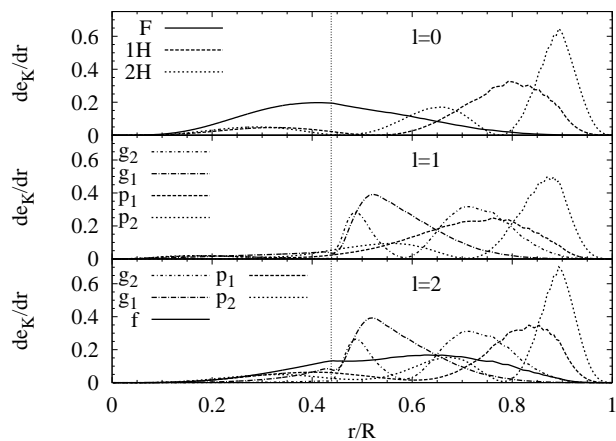}
  \caption{Profiles of $de_{\rm k}/dr \equiv \sigma^2\xi_r^2\rho r^2$, which is proportional to the local kinetic energy density of the oscillations and normalized as $\int_0^R de_{\rm k}/dr\; dr/R=1$, of radial ($l=0$) and non-radial ($l=1,\;2$) modes for the $500M_{\sun}$ model marked with the open circle in Fig. \ref{fig:1}. The dotted vertical line denotes the top of the convective core.}
  \label{fig:3}
\end{figure}

Table \ref{tab:2} shows the results of the stability analysis for the models marked with the open circles in Fig. \ref{fig:1}. The growth time-scale is defined as $\tau_{\rm g}\equiv -2\pi/\sigma_{\rm I}$, that is, the $e$-folding time of the amplitude. The positive values of the growth time-scale mean the instability as well as the growth rate. The radial fundamental mode becomes unstable since the amplitude is relatively large in the core and excitation by the $\varepsilon$-mechanism there exceeds flux dissipation in the envelope. On the other hand, the other radial modes are vibrationally stable since the amplitude is much larger in the envelope than in the core (Fig. \ref{fig:3}). 

Although non-radial modes of $l=1$ and 2 were also analyzed, no unstable modes were found. The acoustic waves of the non-radial p-modes can propagate in a less deep interior than those of the radial oscillations since they are reflected at the depth where the oscillation frequency is equal to the Lamb frequency. The gravity waves of the g-modes cannot propagate in the central nuclear burning region due to the large convective core (Fig. \ref{fig:3}).     

The instability of the fundamental mode continues during the early stage of the core hydrogen burning. The most unstable models, marked with the open circles in Fig. \ref{fig:1}, are at age $\simeq 5\times 10^5$ yr and the growth time-scale is $\sim 10^4$ yr. As pointed out by \cite{Baraffe2001}, the growth time-scale in the case of the $\varepsilon$-mechanism is relatively long. And since the central temperature does not increase efficiently from $500$ to $3000M_{\sun}$, the effect of the nuclear energy generation and the growth rate does either. 

After the activation of the CNO cycle, the stellar radius expands. With this, the density contrast between the inner and outer parts of the star, or the ratio of the central density to the average density in the whole star, $\rho_c/\langle\rho\rangle$,  increases. This causes the amplitude in the envelope becomes much larger than in the core. Fig. \ref{fig:2} shows that in such a stage flux dissipation in the envelope exceeds the excitation by the $\varepsilon$-mechanism, which works efficiently in the core. The instability disappears at age $\simeq 1\times 10^6$ yr, or $X_c\simeq 0.4$. According to \citeauthor{Baraffe2001}, the duration of the unstable phase becomes longer as the stellar mass increases from 120 to 500$M_{\sun}$. In this mass range, the stellar structure may become more favourable to the $\varepsilon$-mechanism instability as the stellar mass increases. In the mass range analyzed in this study, on the other hand, the stellar structure and hence the duration of the unstable phase do not change substantially with stellar mass.

Compared with the result of \citeauthor{Baraffe2001}, the duration of the unstable phase for $500M_{\sun}$ is longer, while the maximum value of the growth rate is lower. In their analysis, around the ZAMS stage, the growth rate initially has a positive and extremely high value, and then shows the rapid variation with time due to the adjustment of the stellar structure when the initial CNO is produced (see their fig. 4). In our analysis, on the other hand, the growth rate is initially negative and becomes positive just after the onset of the CNO cycle. This implies a difference in the methods to calculate evolutionary models.

\subsection{Mass loss}
\cite{Appenzeller1970a,Appenzeller1970b} performed non-linear analysis of vibrational instability of very massive main-sequence stars against radial pulsation. It was demonstrated that recurring shock waves in the outer layers results in the forming of a continuously expanding shell around the star when the velocity amplitude exceeds the sound velocity. Eventually the outermost layers of the shell reach the escape velocity and mass loss occurs. In addition, due to the shock dissipation, the amplitude cannot increase significantly above the value at which the shell is formed. 

\citet{Baraffe2001} performed linear stability analysis of the population III very massive stars and estimated mass loss due to the vibrational instability against the radial fundamental mode. Although the absolute value of the amplitude cannot be determined in the linear analysis, they set the velocity amplitude to the sound velocity at the stellar surface. Besides, they assumed that the pulsation energy gained through the heat-engine mechanism was transformed into the kinetic energy of the escaping matter. 

Although non-linear analysis is actually necessary to investigate the pulsational mass loss, we estimated the mass-loss rate in the same method as \citeauthor{Baraffe2001} The mass loss rate was evaluated for each unstable model and integrated with time. In this study, the equilibrium models evolve without mass loss although their mass should be actually reduced by following the mass loss rate at each evolutionary stage. 

The results are shown in table \ref{tab:3}. $\Delta M$ denotes the total mass lost during the unstable stage. The one for $500M_{\sun}$ is roughly consistent with the result of \citeauthor{Baraffe2001}, $25M_{\sun}$. Besides, in their result, the ratio of the mass-loss amount to the total stellar mass increases from $120$ to $500M_{\sun}$. Extrapolating from their result, significant mass loss was expected to occur for stars with $> 500M_{\sun}$. Such a tendency is also found in our result; 4 per cent for $500M_{\sun}$, while 7.7 per cent for $3000M_{\sun}$. Such mass loss would to some extent cause difference from the evolution without the mass loss, but, by itself, might not have significant influence on the qualitative later scenario of the stars. 

\begin{table}
    \caption{Property of mass loss due to the instability of the fundamental mode. \label{tab:3}}
    \begin{tabular}{ccccc}\hline
      $M$ & $\tau_{\rm uns}$ & $dM/dt$ & $\Delta M$ & $M_{\rm cc}$ \\
      $(M_{\sun})$  & (yr) & $(M_{\sun}\;{\rm yr}^{-1})$  & $(M_{\sun})$ & $(M_{\sun})$ \\ \hline
      500 & $1.2\times 10^6$ & $2.0\times 10^{-5}$ & $2.0\times 10$ & $3.4 \times 10^2$ \\
      1000 & $1.1\times 10^6$ & $6.8\times 10^{-5}$ & $6.1\times 10$ & $6.9 \times 10^2$ \\
      3000 & $1.0\times 10^6$ & $3.1\times 10^{-4}$ & $2.3\times 10^2$ & $2.0 \times 10^3$ \\ \hline
    \end{tabular}

    \medskip
    From left to right the table entries read: stellar mass, duration of vibrationally unstable phase, mass-loss rate for the model marked with the open circles in Fig. \ref{fig:1}, total amount of the mass loss and convective core mass at the evolutionary stage when the instability disappears.
\end{table}

The very massive stars have significantly large convective cores. If the mass loss is so strong that the stellar surface would become close to the top of the convective core, the CNO-elements produced by the triple alpha reaction and the CNO cycle will reach the surface by the convective mixing. This could cause the radiative mass loss to become efficient. For the stars with $500-3000M_{\sun}$, the convective core occupies $\simeq 90$ per cent of the stellar mass at the ZAMS stage (table \ref{tab:1}), but monotonically decreases during the core hydrogen burning stage. Then, the mass fraction of the convective core becomes less than 70 per cent at the evolutionary stage when the instability disappears (table \ref{tab:3}). Thus the amount of the mass loss in our result is not enough to trigger the above situation.  
For the more massive stars, however, the convective core and amount of mass loss are expected to be larger.
Besides, although the instability disappears due to increase in the density contrast as discussed in section \ref{sec:4.1}, the stronger mass loss could maintain the vibrational instability by limiting the density contrast, like models of Wolf-Rayet stars shown in \citet{Maeder1985}. Then, it is worth analyzing if the above situation occurs for stars with $>3000M_{\sun}$. 

\section{Conclusions}
We have analyzed the vibrational stability of the main-sequence stage of population III $500-3000M_{\sun}$ stars against the radial and non-radial modes. We found that only the radial fundamental mode becomes unstable due to the $\varepsilon$-mechanism. The instability appears just after the pre--main-sequence contraction stops and the nuclear energy generation becomes large enough to maintain the energy equilibrium, while it disappears in the middle part of the core hydrogen burning stage as the density contrast between the inner and outer parts of the stars becomes larger. The other radial and non-radial modes never become unstable since the amplitude in the envelope is much larger than in the core. Unfortunately, the result contains uncertainty due to the large convective core. Improvement in our knowledge about the convection-pulsation interaction is required to obtain a definite conclusion on the stability analysis.

To measure the significance of the instability, we estimated the amount of mass loss due to the instability by adopting the \cite{Baraffe2001} method, although non-linear analysis was actually necessary. We found that the amount is less than 10 per cent of the whole stellar mass. Thus, we conclude that mass loss due to this instability, by itself, is not significant for the later evolution. However, we note that \cite{Marigo2003} suggested that the radiative mass loss is supposed to siginificantly affect the evolution of the population III very massive stars with $>750M_{\sun}$. 
\cite{Ekstrom2008} showed that rotational models of population III stars favour the surface enrichment of the heavy elements due to rotational mixing processes, which might trigger stronger radiative mass loss.
Then, the effect of the vibrational instability should be combined with the above effects for more exact investigation of the evolution.

\section*{Acknowledgments}
The authors are grateful to the anonymous referee for his/her useful comments for improving this Letter. The authors are also grateful to K. Omukai for his helpful advices for starting this study. This research has been financed by Global COE Program ``the Physical Sciences Frontier'', MEXT, Japan.

\bsp

\label{lastpage}


\begin{thebibliography}{99}
\bibitem[\protect\citeauthoryear{Abel et al.}{2002}]{Abel2002}
  Abel, T., Bryan, G. L., \& Norman, M. L. 2002, Sci, 295, 93
\bibitem[\protect\citeauthoryear{Appenzeller}{1970a}]{Appenzeller1970a}
  Appenzeller, I. 1970a, A\&A, 5, 355
\bibitem[\protect\citeauthoryear{Appenzeller}{1970b}]{Appenzeller1970b}
  Appenzeller, I. 1970b, A\&A, 9, 216
\bibitem[\protect\citeauthoryear{Bahena \& Klapp}{2010}]{Bahena2010}
  Bahena, D. \& Klapp, J. 2010, Ap\&SS, 327, 219
\bibitem[\protect\citeauthoryear{Baraffe et al.}{2001}]{Baraffe2001}
  Baraffe, I., Heger, A., \& Woosley, S. E. 2001, ApJ, 550, 890
\bibitem[\protect\citeauthoryear{Bromm et al.}{1999}]{Bromm1999} 
  Bromm, V., Coppi, P. S., \& Larson, R. B. 1999, ApJ, 527, L5
\bibitem[\protect\citeauthoryear{Ekstr\"om et al.}{2008}]{Ekstrom2008}
  Ekstr\"om, S., Meynet, G., Chiappini, C., Hirschi, R., \& Maeder, A. 2008, A\&A, 489, 685
\bibitem[\protect\citeauthoryear{Feng \& Soria}{2011}]{Feng2011} 
  Feng, H. \& Soria, R. 2011, New Astron. Rev., 55, 166
\bibitem[\protect\citeauthoryear{Heger et al.}{2003}]{Heger2003} 
  Heger, A., Fryer, C. L., Woosley, S. E., Langer, N. \& Hartmann, D. H. 2003, ApJ, 591, 288 
\bibitem[\protect\citeauthoryear{Klapp}{1983}]{Klapp1983} 
  Klapp, J. 1983, Ap\&SS, 93, 313
\bibitem[\protect\citeauthoryear{Klapp}{1984}]{Klapp1984} 
  Klapp, J. 1984, Ap\&SS, 106, 215
\bibitem[\protect\citeauthoryear{Ledoux}{1941}]{Ledoux1941}
  Ledoux, P. 1941, ApJ, 94, 537
\bibitem[\protect\citeauthoryear{Maeder}{1985}]{Maeder1985}
  Maeder, A. 1985, A\& A, 147, 300
\bibitem[\protect\citeauthoryear{Marigo et al.}{2001}]{Marigo2001}
  Marigo, P., Girardi, L., Chiosi, C., \& Wood, P. R. 2001, A\&A, 371, 152 
\bibitem[\protect\citeauthoryear{Marigo et al.}{2003}]{Marigo2003}
  Marigo, P., Chiosi, C. \& Kudritzki, R. P. 2003, A\&A, 399, 617
\bibitem[\protect\citeauthoryear{Ohkubo et al.}{2006}]{Ohkubo2006}
  Ohkubo, T., Umeda, H., Maeda, K., Nomoto, K., Suzuki, T., Tsuruta, S. \& Rees, M. J. 2006, ApJ, 645, 1352
\bibitem[\protect\citeauthoryear{Omukai \& Palla}{2003}]{Omukai2003}
  Omukai, K. \& Palla, F. 2003, ApJ, 589, 677
\bibitem[\protect\citeauthoryear{Rogers \& Iglesias}{1992}]{Rogers1992} 
  Rogers, F. J., \& Iglesias, C. A. 1992, ApJS, 79, 507 
\bibitem[\protect\citeauthoryear{Schwarzschild \& H\"arm}{1959}]{Schwarzschild1959}
  Schwarzschild, M., \& H\"arm, R. 1959, ApJ, 129, 637
\bibitem[\protect\citeauthoryear{Sonoi \& Shibahashi}{2012}]{Sonoi2012}
  Sonoi, T., \& Shibahashi, H. 2012, PASJ, in press 
\bibitem[\protect\citeauthoryear{Stothers}{1992}]{Stothers1992}
  Stothers, R. B. 1992, ApJ, 392, 706
\bibitem[\protect\citeauthoryear{Umeda \& Nomoto}{2002}]{Umeda2002}
  Umeda, H. \& Nomoto, K. 2002, ApJ, 565, 385
\bibitem[\protect\citeauthoryear{Umeda \& Nomoto}{2005}]{Umeda2005}
  Umeda, H. \& Nomoto, K. 2005, ApJ, 619, 427
\end{thebibliography}
\end{document}